# Standardization of $^{18}$F by Digital β(LS)-γ Coincidence Counting

**Rodrigues D. [1]\*, Balpardo C. [1], Cassette P. [2], Arenillas P. [1], Capoulat M. E. [1], Ceruti G. [1], García-Toraño E. [3]**

[1] *Laboratorio de Metrología de Radioisótopos (LMR), CNEA, Buenos Aires, Argentina.*

[2] *Laboratoire National Henri Becquerel (LNHB), LNE, Paris, France.*

[3] *Laboratorio de Metrología de Radiaciones Ionizantes (LMRI), CIEMAT, Madrid, Spain.*

**Abstract**

The nuclide $^{18}$F disintegrates to $^{18}$O by β$^+$ emission (96.86%) and electron capture (3.14%) with a half-life of 1.8288 h. It is widely used in nuclear medicine for positron emission tomography (PET). Because of its short half-life this nuclide requires the development of fast measuring methods to be standardized. The combination of LSC methods with digital techniques proves to be a good alternative to get low uncertainties for this, and other, short lived nuclides. A radioactive solution of $^{18}$F has been standardized by coincidence counting with a LSC, using the logical sum of double coincidences in a TDCR array and a NaI scintillation detector. The results show good consistency with other techniques like 4πγ and LSC.

**1. Introduction**

The nuclide $^{18}$F disintegrates by β$^+$ emission (96.86%) and electron capture (3.14%) (Figure 1) to $^{18}$O (Bé et al., 2008). Positron annihilation produces two gamma rays of 511 keV and its short half-life makes it suitable for PET practice.

Among the existing alternatives for the standardization of radioactive sources, the coincidence counting method is one of the most widely used in any of its variants (Dunworth, 1940; Campion, 1959; Baerg, 1966). It is based on the simultaneous detection of two different radiations from the same nuclear decay process.

---

\* drodrigu@cae.cnea.gov.ar

The digital coincidence counting (DCC) technique, that appeared in the last years, has significant advantages over the conventional coincidence method, especially in speeding up and simplifying the measurement procedure.

This paper discusses the application of the DCC method to short lived nuclides using a liquid scintillation counter (LSC) in the beta channel. The results are compared with those obtained by the 4πγ method using a well-type NaI(Tl) detector and by direct LSC counting using the logical sum of double coincidence from a three photomultipliers array (TDCR).

## 2. Experimental setup and methods

2.1 *DCC, TDCR and combination of both: 4πβ(LS)-γ*

The basic idea of the DCC technique (Buckman and Ius, 1996) is to measure and register, in a digital way, signals of shaped linear pulses from amplifiers of both beta and gamma channels. Simultaneously, the clock time of the pulses is registered. A data stream of pulse heights and time stamps is sequentially stored in the system memory. After finishing the measurement, the data are transferred to the hard disk of a personal computer as ASCII or binary files.

Recently, the LMR-CNEA has acquired a Time and Amplitude Recorder (TAR) module from ULS-KOREA which records the amplitude and arrival times of the pulses from two independent channels.

To determine the amplitude of the signal, the TAR implements a 16-bit successive approximation ADC (Analog to Digital Converter) in each of its two channels. The pulse information is recorded into a 256MB memory system, type SDRAM (Synchronous Dynamic Random Access Memory), and then, once the measurement is finished, is sent to a PC through the USB port (Park et al., 1998).

The module is controlled by a FPGA (Field Programmable Gate Array), which offers the possibility to modify the parameters of the module by programming the device.

The core of the measurement system is a LSC assembly with three Burle 8850 photomultiplier tubes (PMT), with Ortec 265 bases in a Triple to Double Coincidence array (TDCR) (Pochwalski et al., 1988; Broda and Pochwalski, 1992), which has been previously described by Arenillas and Cassette (Arenillas et al., 2006). All the coincidence logic and imposed dead times are managed by a MAC3 (Bouchard and Cassette, 2000) coincidence module. The amplitude spectrum of each PMT was recorded in order to adjust the counter threshold in each channel. As the correct threshold level was under the specifications of MAC3, a constant fraction discriminator was added between the fast preamplifier and the MAC3 input. This allowed a correct adjustment of the threshold under the single electron peak, for each PMT channel.

To extend this TDCR array to a $4\pi\beta(LS)$-$\gamma$ coincidence setup, a 7.6 x 7.6 cm NaI(Tl) detector was coupled to the optical chamber, in order to measure the gamma emission. And to apply the coincidences method, the logical sum of double coincidences from the TDCR system, were directed to the first TAR channel (playing the role of the beta channel in a regular coincidences system) and the signal from the gamma NaI(Tl) detector, to the second channel. The final system arrangement is showed in Figure 2.

For $^{18}$F, the coincidence equations are not complex due to the simplicity of the decay scheme:

$$N_\beta = N_0 a(\varepsilon_\beta + (1-\varepsilon_\beta)\varepsilon_{\beta\gamma})$$
$$N_\gamma = N_0 2a\varepsilon_\gamma \qquad (1)$$
$$N_c = N_0 2a\varepsilon_\beta \varepsilon_\gamma$$

where $N_0$ is the activity of the source, $N_\beta$ and $N_\gamma$ are the count rates in channel $\beta$ and $\gamma$ respectively, $N_c$ is the coincidence count rate, $a = 0.9686$ (19) is the branching ratio of the $\beta^+$ emission, $\varepsilon_\beta$ and $\varepsilon_\gamma$ are the beta and gamma efficiency respectively and $\varepsilon_{\beta\gamma}$ is the gamma efficiency of the beta detector.

From equations (1) one gets

$$\frac{N_\beta N_\gamma}{N_c} = N_0 a\left(1 + \frac{(1-\varepsilon_\beta)}{\varepsilon_\beta}\varepsilon_{\beta\gamma}\right) \qquad (2)$$

where the contribution of the electron capture branch was neglected because the probability of its emission and detection efficiency (Roteta et al., 2006) is lower than the uncertainties involved. When $\varepsilon_\beta = 1$, the $N_0 a$ value is obtained.

Because of the short half life, the decay during the measurement time has to be taken into account,

$$(N - N_f)\frac{\lambda T e^{\lambda(t_i - t_{ref})}}{1 - e^{-\lambda T}} \qquad (3)$$

where $N$ and $N_f$ are the counting rate and the background in each channel respectively. $T$ is the total time of measurement, $\lambda$ is the decay constant, $t_i$ initial time and $t_{ref}$ the reference time.

A software written in MATLAB (DCCTDCR) was developed at CNEA to perform all calculations, that is, to select gamma gates (511 keV), set dead (50 µs) and resolution (20 µs) times and carry out the counting corrections using the Müller formalism for extendable dead time case (ICRU report, 1994).

## 2.2 $4\pi\gamma$ system

The principle and basis of the $4\pi\gamma$ measurements have been discussed in detail by Winkler, as well as many other authors (Winkler, 1983; Ballaux, 1983) The basic idea is to characterize a well detector in terms of its response to monoenergetic gamma-rays. This can be done either by measuring a set of reference sources (Winkler, 1983) or by a detailed Monte Carlo calculation (García-Toraño et al., 2007). Then, the counting efficiency of any nuclide which emits several gamma rays can be calculated by a combination of the counting efficiency of all decay paths for which a cascade of gamma or X rays is emitted. For a positron emitter in geometry close to $4\pi$, the detection efficiency is given by:

$$\varepsilon = 1 - (1 - \varepsilon_{511})^2 \qquad (4)$$

where $\varepsilon_{511}$ represents the counting efficiency for a gamma ray of 511 keV and the angular correlations between both annihilation photons have been neglected. This expression could only be used if all positrons annihilated in a small region close to the source center.

The system used for these measurements was based on a 17.8 x 17.8 cm NaI(Tl) well detector from BICRON and analog electronics and was characterized in terms of detection efficiency for monoenergetic gamma rays using the PENELOPE code (Salvat et al. 2006). A plot of the counting efficiency as a function of the energy for a point source placed in the well bottom is presented in Figure 3. For $^{18}$F measurements, the detection efficiency was recalculated to take into account the positron flight and the effect of the absorbers, although for this particular case, the differences between values predicted by the curve shown in Figure 3 and that calculated are well below uncertainty calculations and could be neglected.

## 3. Measurements and Results

Two sources were prepared by dropping alicuots of about 9 mg of the mother solution onto 10 ml of Ultima Gold AB in 22 ml glass vials. Each of them was measured during 15 minutes by LSC and 4πβ(LS)-γ method simultaneously.

Also, a single point-like source was prepared by depositing and drying a drop of about 10 mg of the solution between two mylar foils.

One month later, the three vials were measured again looking for impurities, but were not found.

*3.1 Liquid scintillation counting*

Due to positrons' energy (633.5 keV being the maximum value and 249.3 keV the average value) the detection efficiency using LSC is close to one for this branch. In particular, the value obtained for the ratio of triple to double coincidences was 0.994 (1), therefore the efficiency value could be considered equal to unity (Balpardo et al., 2010). Hence, the activity concentration multiplied by the branching ratio was directly calculated as the ratio between the logical sum of double coincidences and the live measurement time.

Two 15-minute-long measurements were performed, and the average activity concentration obtained was 2970 (16) kBq.g$^{-1}$.

*3.2  4πβ(LS)-γ method*

When the coincidence method is not based on LS, extrapolation is necessary (Roteta et al., 2006), even for $^{18}$F. In our case, for the same reasons quoted before, the beta efficiency was taken equal to one and the activity concentration $a_c$ was calculated using equation 5, i.e. without extrapolation.

$$a_c = \frac{N_\beta \cdot N_\gamma}{a \cdot N_c \cdot m} \quad (5)$$

The same two measurements were analyzed by the coincidence method, and the average obtained was 2970 (17) kBq.g$^{-1}$.

### 3.3  4πγ system

The point-like source was placed between two aluminum discs 2.40 mm thick in order to obtain a point-like gamma source (most positrons annihilate in a short distance).

Five 1000-second-long measurements were performed and the average obtained was 2978 (18) kBq.g$^{-1}$. Figure 4 shows a comparison of the results obtained by the three methods.

### 3.4 Final result and uncertainty

Taking into account that the results from the various methods are correlated, mainly due to weighing and positron emission probability, we first calculate the weighted mean value (JCGM 100:2008.GUM) of the results obtained by *LCS* and *4πβ(LS)-γ* methods and obtain:

2970 (16) kBq. g$^{-1}$

Here the uncertainties (Table I) from weighing and positron emission probability are considered separately, based on the fact that uncertainties are correlated due to the use of the same sources.

The weighted average between this result and the activity concentration from the *4πγ* method, provides the final result:

2974 (12) kBq. g$^{-1}$

where the uncertainty from the positron emission probability is considered separately for the same reason as above.

All results are included within a region corresponding to a coverage factor *k*=1.

Moreover, the differences between results obtained by 4πγ, LSC and 4πβ(LS)-γ methods and the average value do not exceed 0.15 %.

## 4. Discussion and Conclusions

A setup of digital coincidence counting using a liquid scintillation counter as beta channel has been used in the standardization of $^{18}$F, and the results obtained by this technique are consistent with those obtained by LSC and 4πγ counting. The easiness of source preparation in LSC and the versatility of the digital method combine here to provide an excellent alternative to standardize many nuclides, and makes it especially convenient for short lived nuclides.

Although it has not been the main goal of this work, the system is capable of implementing simultaneously the TDCR and coincidence methods. In this case, it will be necessary to vary the efficiency by using optical filters and to calculate the system efficiency with specific program (Rodrigues et al. 2008).

In particular, for the 4πβ(LS)-γ arrangement, taking advantage of the digital recording process and the off-line analysis, it will be possible to apply several approaches for one measurement: coincidence, anticoincidence, correlation and coincidence with bi-dimensional extrapolation (Bobin, 2007).

## 5. Acknowledgements

The authors would like to express their thanks to Dr. Miguel Roteta, Dr. Virginia Peyres and Lic. Roberto Llovera for constant attention to this work and useful discussion.

|  | Method | | |
| --- | --- | --- | --- |
| Source of uncertainty | DCC | LSC | 4πγ |
| Counting statistics | 0.241 | 0.057 | 0.100 |
| Half-life | 0.030 | 0.075 | 0.334 |
| Dead time | 0.003 | <<$10^{-3}$ | 0.100 |
| Background | <<$10^{-3}$ | <<$10^{-3}$ | <<$10^{-3}$ |
| Resolution time | <<$10^{-3}$ | --- | --- |
| Delay time | <<$10^{-3}$ | --- | --- |
| Weighing | 0.482 | | 0.394 |
| Branching ratio | 0.196 | | |
|  |  |  |  |
| Relative combined uncertainty (%) | 0.57 | 0.53 | 0.57 |

Table I. Relative uncertainty components (in %) in the determination of the activity concentration measured with three methods.

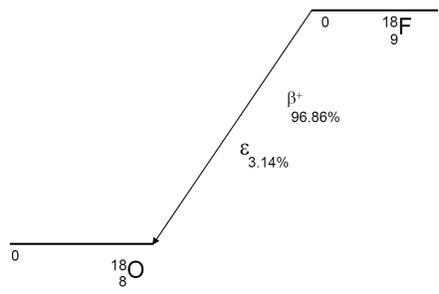

Figure 1

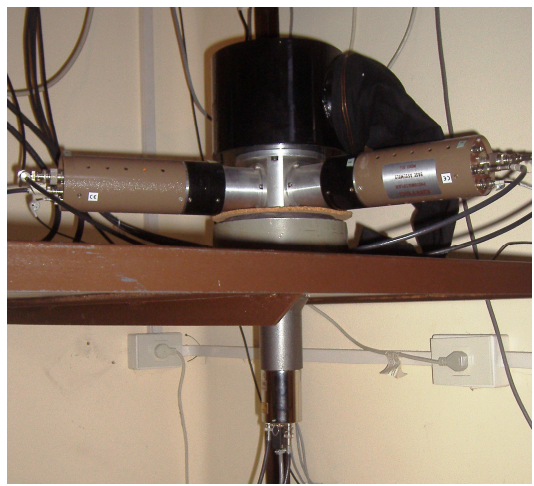

Figure 2

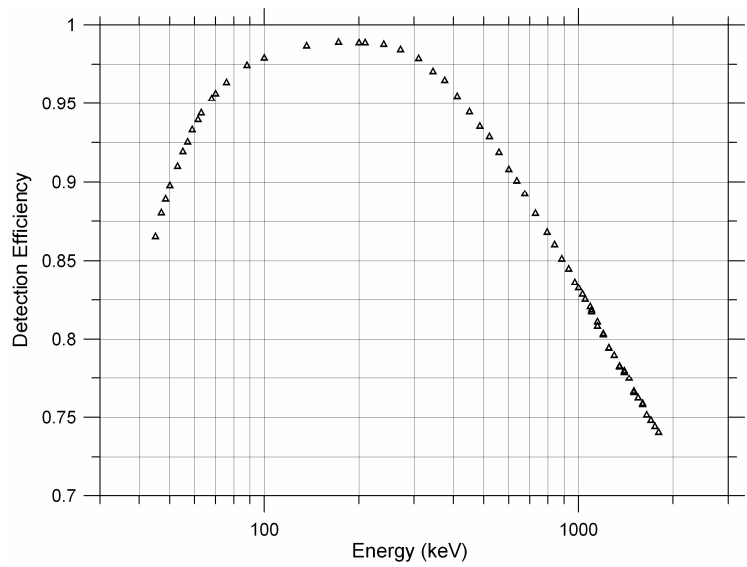

Figure 3

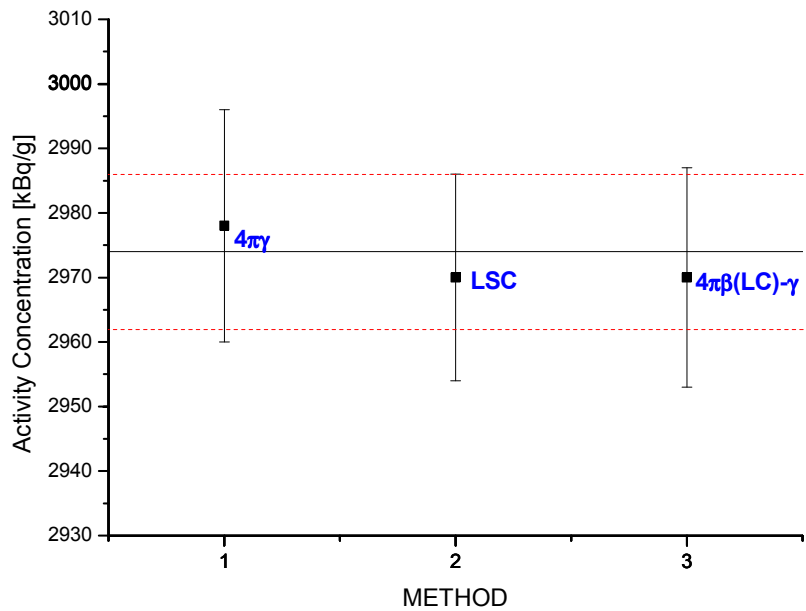

Figure 4

**Figure Captions**

Figure 1. Decay scheme of $^{18}$F. Data taken from Bé et al. (2008).

Figure 2. $4\pi\beta$(LS)-$\gamma$ coincidence setup.

Figure 3. Detection efficiency for a 17.8 x 17.8 cm. NaI(Tl) well detector as a function of the gamma ray energy. The combined efficiency for the positron-emitter branch of $^{18}$F is close to unity.

Figure 4. Results obtained in the standardization of $^{18}$F by the methods described in this paper.